\documentstyle[12pt,epsfig,amsmath,amssymb,psfrag,pifont,cite]{article}
\numberwithin{equation}{section}
\numberwithin{table}{section}
\newcommand{\lambdabf}{{\mbox{\boldmath $\lambda$}}}

\renewcommand{\d}{\partial}

\def\ga{\mathrel{\raise.3ex\hbox{$>$\kern-.75em\lower1ex\hbox{$\sim$}}}}
\def\la{\mathrel{\raise.3ex\hbox{$<$\kern-.75em\lower1ex\hbox{$\sim$}}}}

\def\I_M{{I_{\scriptscriptstyle M\times M}}}

\def\be{\begin{equation}}
\def\ee{\end{equation}}
\def\bea{\begin{eqnarray}}
\def\eea{\end{eqnarray}}

\newcommand{\beqal}{\begin{eqnarray}\label}
\newcommand{\beqa}{\begin{eqnarray}}
\newcommand{\eeqa}{\end{eqnarray}}

\begin{document}

\begin{titlepage}
\begin{center}

\vskip .2in

{\Large \bf Universal properties of thermal and electrical conductivity of gauge theory plasma from holography}
\vskip .5in

{\bf Sachin Jain\footnote{e-mail: sachjain@iopb.res.in}\\
\vskip .1in
{\em Institute of Physics,\\
Bhubaneswar 751~005, India.}}
\end{center}\noindent
\baselineskip 15pt

\begin{center} {\bf ABSTRACT}

\end{center}
\begin{quotation}\noindent
\baselineskip 15pt
We propose that for conformal field theories admitting gravity duals, the thermal conductivity is fixed by the central charges in a universal manner. Though we do not have a proof as yet, we have checked our proposal against several examples. This proposal, if correct, allows us to express electrical conductivity in terms of thermodynamical quantities even in the presence of chemical potential.

\end{quotation}
\vskip 2in
March 2010\\
\end{titlepage}
\vfill
\eject

\setcounter{footnote}{1}
\section{Introduction}
The gauge/gravity duality allows us to 
gain insights into various properties of strongly coupled gauge theories both at zero and non-zero
temperature. In particular, the transport coefficients of strongly coupled gauge theories which are hard to compute otherwise, can be computed easily using gauge/gravity duality.  Even though these theories, in several ways, are different from theories such as QCD, they do share qualitatively similar behavior. For instance, elliptic flow measurements at RHIC (see \cite{Teaney:2009zz} and references therein for details ) seem to indicate a very small  ratio $\frac{\eta}{s}\lesssim 0.3 $, which implies that QGP behaves like a nearly ideal fluid \footnote{ Let us note that this ratio for all the known material have much larger value (see \cite{Kovtun:2004de} for details), for example for water, liquid nitrogen etc $\frac{\eta}{s}\gg 1$ and for liquid helium $\frac{\eta}{s}\approx 0.8 $, in the dimension less units.}. Moreover calculation using AdS/CFT shows universal ratio
 \begin{equation}
  \frac{\eta}{s} = \frac{1}{4 \pi} \approx 0.08,\label{etabys}
 \end{equation} which falls into the experimental range observed at RHIC.
Motivated by this there are many papers which explores the  universal behavior of other transport coefficients \cite{Policastro:2001yc}.
     
In this note we conjecture that the thermal conductivity also shows some universal behavior. More precisely, we propose that for a $d$ dimensional strongly coupled gauge theory
 \begin{equation}
  \frac{\kappa_T}{\eta T}\sum\limits_{j=1}^m(\mu^{j})^{2} = \frac{d^2}{d-2}\Big(\frac{c^{'}}{k^{'}}\Big) = 8 \pi^{2}\frac{d-1}{d^3 (d+1)}\frac{c}{k}\label{unvsalthrmalcon1},
 \end{equation}  where $\kappa_T$ is the thermal conductivity (the heat current response to thermal gradient in the absence of electrical current), $T$ is the temperature, $\mu$ the chemical potential, $\eta$ the shear viscosity and $c, k$ are central charges of dual gauge theory. We test our proposal against several examples. However a general proof of this result is still lacking.

In the next section, after reviewing the standard result for viscosity to conductivity ratio at vanishing chemical potential,  we show that eqn. (\ref{unvsalthrmalcon1}) holds at $\mu =0$.\footnote{Let us note that in the presence of equal number of positive and negative charges, chemical potential is zero.} Then, based on few examples, we conjecture that eqn. (\ref{unvsalthrmalcon1})  holds true even for arbitrary nonzero chemical potential. Subsequently, using eqn. (\ref{unvsalthrmalcon1}), we show a way to compute electrical conductivity in terms of thermodynamical quantities alone even in the presence of non-zero chemical potential. Conclusions and discussions are given in section 4. A brief discussion on electrical and thermal conductivity is presented in appendix A. In appendix B, we provide necessary details of the results that are used in the text.
\section{Thermal and electrical conductivity}
 In this section we first review the relation between electrical conductivity and shear viscosity  at vanishing chemical potential \cite{Kovtun:2008kx}.
 In a CFT, short distance physics is described by singularities of correlation functions  where central charges of the theory appear explicitly (in this energy scale effects of temperature are irrelevant). For example let us consider correlation functions of energy momentum tensor $T_{\mu\nu}$ and $U(1)$ conserved current $J_{\mu}$
\begin{equation}
 \langle J(x)J(0)\rangle\sim \frac{k}{x^{2(d-1)}}, ~~~~ \langle T(x)T(0)\rangle\sim \frac{c}{x^{2 d}}, 
\end{equation} where central charges $c, k$ measure the number of total degrees of freedom and the number of charged degree of freedom of the system\footnote{ So we expect $k \leq c $.} respectively. 
We also know that at long distances physics is described by thermodynamics and transport coefficients. In this scale, the effect of temperature becomes important.  To describe equilibrium at $T\neq 0$, we look at pressure and charge susceptibility $\chi = \frac{\rho}{\mu} $ where $\rho(T, \mu)$ is the charge density . If $T$ is the only scale in the theory \footnote{to define $\chi$, one can introduce small chemical potential and see the effect in $\rho$ .}, then 
 \begin{equation}
 P = c^{'}T^{d} ,~~~~~~~~~~~~~~~~~ \chi = k^{'} T^{d-2},\label{thermo}
\end{equation} where $c^{'}, k^{'}$ measure the number of total degree of freedom and number of charged degree of freedom at that scale. For $d>2$, in general there is no relation between $c, c^{'}$ and $k, k^{'}$. But it was shown in \cite{Kovtun:2008kx} that for CFT's which admit gravity duals, there exist such relation and are given by
\begin{equation}
\frac{c^{'}}{c} = \frac{1}{4 \pi^{\frac{d}{2}}}\Big(\frac{4 \pi}{d}\Big)^{d}\frac{\Gamma(d/2)^{3}}{\Gamma(d)} \frac{d-1}{d(d-1)},~~~~~~  \frac{k^{'}}{k} = \frac{1}{2 \pi^{\frac{d}{2}}}\Big(\frac{4 \pi}{d}\Big)^{d-2}\frac{\Gamma(d/2)^{3}}{\Gamma(d)} \label{ck}
\end{equation} where\footnote{In our notation $d$ is the dimension of gauge theory.}$ d \geq 3$. 

It is well known that, in this class of CFT's, even certain transport coefficients are determined in-terms of thermodynamical quantities  (for example $\eta = \frac{s}{4 \pi} $). Other such relation between viscosity and conductivity ($\sigma$) at vanishing chemical potential ($\mu = 0 $) is
\begin{equation}
\frac{\eta}{\sigma T^2} = (d-2) \Big(\frac{c^{'}}{k^{'}}\Big) = 8 \pi^{2}\frac{(d-1)}{(d-2)d (d+1)}\frac{c}{k}.\label{unvsalelectricalcon}
\end{equation}  Eqn. (\ref{unvsalelectricalcon}) implies at vanishing chemical potential i.e. at $\mu = 0 $, electrical conductivity can be computed in terms of central charges only. Using  eqn. (\ref{unvsalelectricalcon}),  (\ref{thermo}) and $s = d~ c^{'}T^{d-1} $ one gets 
\begin{equation}
  \eta  = \frac{d}{4 \pi} c^{'}T^{d-1}, ~~~~~~\sigma=\frac{1}{d-2}\frac{d}{4 \pi}k^{'} T^{d-3} \label{trnsport}.
 \end{equation} 
Since thermodynamics is determined by the central charges,  we conclude  that the momentum ($\eta$) and charge ($\sigma$) transport are fixed by thermodynamics\footnote{As an aside lets review membrane paradigm arguments. 
It was shown in \cite{Iqbal:2008by} using membrane paradigm arguments that at $\mu = 0 $, electrical conductivity can be determined in terms of geometry only. 
If we use the results in  \cite{Iqbal:2008by}, we immediately reach at
\begin{equation}
\frac{\eta}{\sigma T^2} = \frac{1}{T^2}\frac{g_{d+1}^2}{16 \pi G_N} g_{xx}(r_0).
\end{equation}
As an example consider a CFT with the gravity dual given by $AdS_{d+1}$ with $d\neq 3$, which has a metric
\begin{equation}
ds^{2}= \frac{r^2}{R^2} \Big(- f(r) dt^{2} + dx_1^2 + \cdots + dx_{d-1}^2 \Big) + \frac{R^2}{f(r) r^2} dr^2,
\end{equation}
 with $ f(r)= 1 - (\frac{r_0}{r})^{d} $ and hawking temperature is $T_{H} = \frac{d}{4 \pi }\frac{r_0}{R^2}$ where $r_0$ and R are the position of horizon and AdS curvature scale respectively. Using the above relations we obtain, 
\begin{equation}
\frac{\eta}{\sigma T^2} = \frac{\pi}{d^2}\frac{R^2 g_{d+1}^{2}}{G_{d+1}}
\end{equation} which is same as reported in \cite{Kovtun:2008kx}.}. Existence of such relations between thermodynamics and transport coefficient are interesting\footnote{We note that hydrodynamics description is valid in the energy range $\omega \ll T$ which is collision dominated regime \cite{Damle}}, since transport coefficients are characterized by inelastic collisions among thermally excited carriers (of energy $\sim T$) hence they are not fixed in terms of thermodynamics. 

~~~~~~~~~~~~~~What we conclude from above discussion is that, at non zero temperature and at $\mu = 0$, certain transport coefficients are determined by thermodynamics. It is interesting to ask whether for $\mu \neq 0 $ and at finite temperature, transport coefficients can be determined from thermodynamics. We note that in this case it is already known that (\ref{etabys}) still holds i.e. momentum transport can be determined solely by thermodynamics. It would be interesting if one can express the electrical conductivity which encodes the charge transport, in terms of thermodynamics. 

We now proceed  to provide  evidences in favor of equation (\ref{unvsalthrmalcon1}). In what follows, we first derive equation for $\mu =0$ and then provide support for cases with $\mu \neq 0$.

\begin{itemize}
 \item \textbf{Derivation of Eqn. (\ref{unvsalthrmalcon1}) for $\mu = 0$ :}
Let us consider theory at small (single) chemical potential and consider the ratio $\frac{\kappa_T}{\eta T} \mu^{2}$. Using the relation (see \cite{Son:2006em} and the Appendix for details) $\kappa_{T} = \Big(\frac{\epsilon + P}{\rho}\Big)^2 \frac{\sigma}{T}$, one obtains
\begin{equation}
\frac{\kappa_T}{\eta T} \mu^{2} = \Big( \epsilon + P \Big)^{2} \frac{1}{\Big(\frac{\rho}{\mu}\Big)^{2}} \frac{1}{\Big(\frac{\eta}{\sigma T^2}\Big)}\frac{1}{T^4}.
\end{equation}
Now taking $\mu\to 0$, using $\epsilon = (d-1) P $, $\chi = \frac{\rho}{\mu}$   we immediately get
 \begin{equation}
  \frac{\kappa_T}{\eta T} \mu^{2} = \frac{d^2}{d-2}\Big(\frac{c^{'}}{k^{'}}\Big) = 8 \pi^{2}\frac{d-1}{d^3 (d+1)}\frac{c}{k}.\label{unvsalthrmalcon2}
 \end{equation}
 \end{itemize}

\begin{itemize}
 \item \textbf{Support for Eqn. (\ref{unvsalthrmalcon1}) for $ \mu \neq 0$ :}  For non zero chemical potential, we recall some of the results already reported in the literature. In each case we show that they follow eqn. (\ref{unvsalthrmalcon1}).  Here we tabulate the results for strongly coupled gauge theories having gravity duals ( in the appendix we provide all the required details about gravity duals of these theories ) in the presence of single non zero chemical potential \cite{Son:2006em, Hartnoll:2009sz,Ge:2008ak, Jain:2009pw}.

\begin{center}
\begin{tabular}{l*{2}{c}r}
 Gravity theory in $d+1$ dimension  ~~~~~~~~~~~~~~~~~~~~~       & $\frac{\kappa_T\mu^{2}}{\eta T}$ ~~~~~~~~~~~~ &$\frac{d^2}{d-2}\Big(\frac{c^{'}}{k^{'}}\Big)$ \\
\hline
 R-charge B.H. in $4+1$ dim.           &$8 \pi^{2}$&$8 \pi^{2}$ \\
 R-charge B.H. in $3+1$ dim.          &$32 \pi^{2}$&$32 \pi^{2}$ \\
 R-charge B.H. in $6+1$ dim.          &$2 \pi^{2}$&$2 \pi^{2}$ \\
 Reissner-Nordstrom B.H. in $3+1$ dim. &$4 \pi^{2}\gamma^{2} $ &$4 \pi^{2}\gamma^{2} $\\
 $D3/D7$ in $3+1$ dim.  &$2  \pi^{2}\frac{N_{c}}{N_{f}}$ &$ 2  \pi^{2}\frac{N_{c}}{N_{f}}$\\
\end{tabular}             \end{center}

It was further reported in \cite{Jain:2009pw} that for the R-charged black holes in five, four and seven 
dimensions the appropriate ratio of thermal conductivity and viscosity, regardless
of 
the number of charge contents,  are $8 \pi^2$, $32 \pi^2$ and $2 \pi^2$ 
respectively.  Based on these observations we propose that, even in the presence of  finite chemical potential (and arbitrary number of them) we can write
\begin{equation}
  \frac{\kappa_T}{\eta T}\sum\limits_{j=1}^m(\mu^{j})^{2} = \frac{d^2}{d-2}\Big(\frac{c^{'}}{k^{'}}\Big) = 8 \pi^{2}\frac{d-1}{d^3 (d+1)}\frac{c}{k}\label{unvsalthrmalcon}.
 \end{equation}
In the next section we use (\ref{unvsalthrmalcon}) to express electric conductivity in terms of the thermodynamical quantities alone.
\end{itemize}
\section{Electrical conductivity } Let us first write down various expressions for thermodynamical quantities, transport coefficients such as viscosity, electrical conductivity in the presence of single chemical potential. In our definition, $\chi = \frac{\rho}{\mu}$. In case of nonzero chemical potential we expect different thermodynamical quantities and transport coefficients to get modified from that of eqn. (\ref{thermo}), (\ref{trnsport})  . In general these can be written as
\begin{equation}
P = c^{'}T^{d}f_{p}(m),~~~~~~~~~~~~~~~~~ \chi = k^{'} T^{d-2}f_{\chi}(m),\label{P}
\end{equation} and
 \begin{equation}
\sigma =\frac{1}{d-2}\frac{d}{4 \pi}k^{'} T^{d-3} f_{\sigma}(m),~~~~~~~~~~~~~~~~~\eta = \frac{d}{4 \pi} c^{'}T^{d-1}f_{\eta}(m),\label{et}
\end{equation} 
where $ m=\frac{\mu}{T}$ and $f(m)$'s are defined such that  $f(m=0)=1$. Now using 
\begin{equation}
\frac{\kappa_T\mu^{2}}{\eta T} = \left( \frac{(\epsilon+P) \mu}{ \rho }\right)^2\frac{\sigma}{\eta T^2}=\frac{d^2}{d-2}\Big(\frac{c^{'}}{k^{'}}\Big),
\end{equation} we get an important constraint between the function $f(m)$'s
\begin{equation}
 \frac{f_{p}^2 f_{\sigma}}{f_{\chi}^{2} f_{\eta}} =1,\label{constraint}
\end{equation} which gives $ f_{\sigma} = \frac{f_{\chi}^{2} f_{\eta}}{f_{p}^2}$.
 We then obtain expression for conductivity\footnote{ We may also write it as, $ \sigma =\frac{d^2}{d-2}\Big(\frac{c^{'}}{k^{'}}\Big) \chi^{2} \frac{\eta  T^2 }{(\epsilon+P)^{2}}  $ , where $\chi = \frac{\rho}{\mu}$.}
\begin{equation}
\sigma =\frac{1}{d-2}\frac{d}{4 \pi}k^{'} T^{d-3} \frac{f_{\chi}^{2} f_{\eta}}{f_{p}^2} \label{univconduct1},
\end{equation} which is entirely fixed in terms of central charges (and thermodynamic quantities).
 For illustrative purpose we end this section with a specific example. Let us consider gauge theory dual to single R-charged five dimensional blackhole.

Using (\ref{univconduct1}) and $f_{\chi},f_{\eta},f_{p}$ and $ k^{'}  $ written in the appendix $ (\textbf{B}) $, equation (B.$20$)  we obtain 
\begin{equation}
 \sigma = N^2 T\Big(\frac{2+ \kappa_1}{ 32 \pi}\Big),
\end{equation} where $\kappa_1$ can be expressed in terms of $m$. This is same as the result reported in the literature \cite{Maeda:2008hn,Jain:2009pw}.

\section{Discussion}
 We have conjectured that, for strongly coupled gauge theories admitting gravity dual, there exists a universal relation between the thermal conductivity and viscosity. This implies that the thermal conductivity can be fixed in terms of the central charges (thermodynamics) alone. We provided several examples satisfying our conjecture. Further more, using our proposed relation, we showed how electrical conductivity can be fixed in terms of thermodynamics even for $\mu \neq 0$.

In general we have $ c > k$, i.e. the charge degree of freedom is less than the total degree of freedom. As a consequence we get a bound
\begin{equation}
\frac{d^3 (d+1)}{8 \pi^{2}(d-1)} \frac{\kappa_T}{\eta T}\sum\limits_{j=1}^m(\mu^{j})^{2} = \frac{c}{k}\geq 1.
\end{equation} 
We can also examine how the bound on $\frac{\sigma}{\chi}$ gets modified in the presence of a chemical potential.
To prove the relation stated in eqn. (\ref{unvsalthrmalcon}), we suspect that the  membrane paradigm arguments might be helpful (It is quite similar to ratio $\frac{\eta}{s}$, which remains same for vanishing and nonvanishing chemical potential). It will also be interesting to see whether the other transport coefficients such as thermo-electric coefficient can be be fixed using results reported in this paper. Also it is of interest to explore as to how eqn. (\ref{unvsalthrmalcon}) as well as the bound get modified when  higher derivative corrections are taken into account. These are currently under investigation.

\section*{Acknowledgments}
It is pleasure to thank Somen Bhattacharjee, Balram Rai, Anirban Basu for extremely useful discussions and Sayan Chakrabarti for patiently listening to the results more than once. I am also thankful to Sayan Chakrabarti, Souvik Banerjee for comments about improving the manuscript and to referee for pointing out required improvements in the presentation. Special thanks to Sudipta Mukherji for taking personal care for improving the presentation of the text. I would also like to thank Sudipta Mukherji , Binata Panda, Souvik Banerjee and in particular Sankhadeep Chakraborty for encouragement.
\renewcommand{\thesection}{\Alph{section}} 
\setcounter{section}{0} 

\section{Electrical and Thermal conductivity:}  In this appendix we present, in brief, the definitions of  electrical and the thermal conductivities that are discussed in the text. For more detail,  see \cite{Son:2006em,Hartnoll:2009sz,Ge:2008ak,Jain:2009pw,Maeda:2008hn,Jain:2009uj}.
\begin{itemize}
 \item {\textbf{Electrical conductivity:}} The electrical conductivity is usually computed from current-current correlator 
\begin{equation}
 \lambdabf = - \lim_{\omega \rightarrow 0} \frac{G_{xx}(\omega, q=0)}{i \omega} = \lim_{\omega \rightarrow 0}\frac{1}{2 \omega}\int_{- \infty}^{\infty}dt~~e^{-i \omega t}\int d\vec{x}\langle[J_{x}(t,\vec{x}), J_{x}(0,\vec{0})]\rangle .
\end{equation} Upon computing one finds 
\begin{equation}
 \lambdabf = -\frac{\rho^{2}}{\epsilon + P}\frac{i}{\omega} + \sigma,
\end{equation} where $\rho$, $\epsilon$ and $P$ are the charge density, energy density and pressure of the fluid respectively and
\begin{equation}
 \sigma = \Re (\lambdabf) = -\lim_{\omega \rightarrow 0} \frac{\Im[G_{xx}(\omega, q=0)]}{ \omega}. 
\end{equation} Let us note that in the presence of charge density, in the limit $\omega \rightarrow 0$, $\Im(\lambdabf) = -\frac{\rho^{2}}{\epsilon + P}\frac{1}{\omega}$ diverges, where as $\sigma$ remains finite. In the text the conductivity that we have discussed is $\sigma$.

\end{itemize}
\begin{itemize}
\item {\textbf{Thermal conductivity:}}
In the following we first review the hydrodynamics with multiple conserved charges (see \cite{Jain:2009pw}) and write down an expression for thermal conductivity. The single charge case was discussed in  \cite{Son:2006em} .
\end{itemize}
\begin{itemize}
 \item \textbf{\textit{Relativistic hydrodynamics with multiple conserved charge:}}
 The continuity equations are normally presented as
\begin{equation}
   \d_\mu T^{\mu\nu}=0, ~~~ \d_\mu J_{i}^\mu =0
\end{equation}
where 
\begin{equation}
   T^{\mu\nu} = (\epsilon+P) u^\mu u^\nu + P g^{\mu\nu}+\tau^{\mu\nu}
    ,~~~
    J_{i}^\mu = \rho_{i} u^\mu + \nu_{i}^\mu\label{velo}
\end{equation}
In the above $\epsilon$ and P are the local energy density and pressure respectively, $u^\mu$ is the local velocity and it obeys $ u_\mu u^\mu = -1$, where as $\tau^{\mu\nu}$ and $\nu_{i}^\mu$ are the dissipative parts of stress-energy tensor and current.

 One can choose
 $u^\mu$ and $\rho_{i}$'s so that 
\begin{equation}\label{transverse}
    u_\mu \tau^{\mu\nu} = u_\mu \nu^\mu_{i} = 0\,.
\end{equation}
Note that
\begin{equation}\label{udT}
    u_\nu \d_\mu T^{\mu\nu} = - (\epsilon+P)\d_\mu u^\mu
    - u^\mu \d_\mu \epsilon + u_\nu \d_\mu \tau^{\mu\nu} =0.
\end{equation}
We also have 
\begin{equation}
    \epsilon+P = Ts +\sum\limits_{i=1}^m \mu^{i} \rho_{i}, \qquad d\epsilon = Tds +\sum\limits_{i=1}^m \mu^{i} d\rho_{i}.\label{thermo1}
\end{equation}
Using which one gets
\begin{equation}
    \d_\mu \left(su^\mu - \sum\limits_{i=1}^m\frac{\mu^{i}}{ T}\nu^\mu_{i}\right) =
    -\sum\limits_{i=1}^m\nu^\mu_{i} \d_\mu \frac{\mu^{i}}{T} - \frac{\tau^{\mu\nu}}{T} \d_\mu u^\mu\,.
\end{equation}
Now $\d_\mu \left(su^\mu - \sum\limits_{i=1}^m\frac{\mu^{i}}{ T}\nu^\mu_{i}\right)$ can be interpreted as the divergence of the entropy current which implies right hand side is positive. So we write
\begin{equation}
 \nu^{\mu}_i = -\sum\limits_{j=1}^m\varkappa_{ij} \left(\d^\mu\frac{\mu^{j}}{ T}+ u^\mu u^\lambda \d_\lambda \frac{\mu^{j}}{ T}\right)
\end{equation} and similarly for $\tau^{\mu\nu}$ (see \cite{Son:2006em}) .
 To interpret $ \varkappa_{ij} $  as the
 coefficient of thermal conductivity, consider no charge current i.e. $ J^\alpha_{j} = 0, $\footnote{In our notation $\mu ,\nu$ runs from $t,1,2...D$ , where as $\alpha$ runs from $1,2,...D$, and i,j are R-charge indices.}  but there is an energy flow,
  $ T^{t\alpha}\neq 0 $, which is the heat flow.  Take $ u^\alpha $ to be small so that one gets using eqn. (\ref{velo}) 
\begin{equation}
 \rho_{i} u^\alpha=\sum\limits_{j=1}^m\varkappa_{ij} \d^\alpha\frac{\mu^{j}}{T}.\nonumber\\
\end{equation}

From which one can obtain
\begin{equation}
 \sum\limits_{i,j=1}^m\rho_{i}\varkappa_{ij}^{-1} \rho_{j} u^\alpha = \sum\limits_{i=1}^m\rho_{i}\d^\alpha \frac{\mu^{i}}{T}\,,
\end{equation}
hence 
\begin{equation}
  u^\alpha =\frac{1}{\sum\limits_{i,j=1}^m\rho_{i}\varkappa_{ij}^{-1}\rho_{j}} \sum\limits_{l=1}^m\rho_{l}\d^\alpha \frac{\mu^{l}}{T}\,.
\end{equation}
 Using eqn. (\ref{thermo1}) we get
\begin{equation}
\sum\limits_{i=1}^m\rho_{i}\d^\alpha \frac{\mu^{i}}{T}=-\frac{\epsilon+P}{T^{2}}\d^\alpha T+\frac{\d^\alpha P}{T}
\end{equation}
after substitution, this gives
\begin{equation}
  u^\alpha =-\frac{1}{\sum\limits_{i,j=1}^m\rho_{i}\varkappa_{ij}^{-1}\rho_{j}}(\frac{\epsilon+P}{T^{2}})(\d^\alpha T-\frac{T}{\epsilon+P}\d^\alpha P) \,.
\end{equation}
 Therefore
  \begin{equation}
     T^{t\alpha} = (\epsilon+P) u^\alpha = -\frac{1}{\sum\limits_{i,j=1}^m\rho_{i}\varkappa_{ij}^{-1}\rho_{j}}(\frac{\epsilon+P}{T})^{2}(\d^\alpha T-\frac{T}{\epsilon+P}\d^\alpha P)
 \end{equation}
hence the coefficient of thermal conductivity can be read off as
 \begin{equation}
   \kappa_T =\left( \frac{\epsilon+P}{ T}\right)^2\frac{1}{\sum\limits_{i,j=1}^m\rho_{i}\varkappa_{ij}^{-1}\rho_{j}}\label{thermalconductivity}.
 \end{equation}
Note that, $\varkappa_{ij}$ can be found out from greens function as 
\begin{equation}
   G_{ij}^{xx} (\omega, q=0) = -i \omega \frac{\varkappa_{ij}}{ T}\ = -i \omega \sigma_{ij},
 \end{equation}
where $ J^{x}_{i}=-G_{ij}(\omega,q=0)A^{x}_j $  and $ \sigma_{ij}(\omega,q=0)$ can be obtained using current-current correlator as discussed earlier.

\end{itemize}

Let us note that for single charge black hole $\frac{1}{\rho_{i}\varkappa_{ij}^{-1}\rho_{j}}=\frac{\varkappa}{\rho^{2}}$. Therefore
\begin{equation}
   \kappa_T =\left( \frac{\epsilon+P}{ \rho T}\right)^2\varkappa=\left( \frac{\epsilon+P}{ \rho }\right)^2\frac{\sigma}{T}.\label{themalcon}
 \end{equation}

\section{Examples} Here we present computations which led to the results of Table in section 2.
\begin{itemize}
 \item \textbf{AdS$_4$ Reissner-Nordstrom blackhole:}
The action is 
 \begin{equation}
  S = \int d^{4}x \sqrt{-g}\Big[ \frac{1}{2 \kappa^2}(R + \frac{6}{L^2})-\frac{1}{4 g^2} F^{2}\Big].
 \end{equation}
Metric is given by (for details see \cite{Hartnoll:2009sz})
\begin{equation}
 ds^{2} = \frac{L^2}{r^2}(-f(r) dt^{2} + \frac{dr^{2}}{f(r)}+ dx^{i}dx^{i}).
\end{equation}
Thermodynamical quantities are given by
\begin{equation}
 T = \frac{1}{4\pi r_{+}}(3-\frac{r^{2}_{+} \mu^{2}}{\gamma^{2}}),~~ P = \frac{L^2}{2\kappa^{2}r^{3}_{+}}(1+\frac{r^{2}_{+} \mu^{2}}{\gamma^{2}})
\end{equation} and
\begin{equation}
S = \frac{2\pi}{\kappa^{2}}\frac{L^{2}}{r^{2}_{+}}, ~~~  \chi = \frac{\rho}{\mu} = \frac{2 L^2}{\kappa^2}\frac{1}{r_{+}\gamma^{2}}
\end{equation} where $r_{+}$ is the horizon radius and $\gamma^2 =\frac{ 2 g^2 L^2}{\kappa^2}$. 
To find out $c^{'}$ and $k^{'}$ best is to set $\mu$ to zero (then express $r_{+}$ in terms of T) and compare with eqn.(\ref{thermo}). After doing this one finds
\begin{equation}
 c^{'}= \frac{ L^2}{2\kappa^2}(\frac{4\pi}{3})^{3}, ~~ k^{'} = \frac{8 \pi}{3}\frac{ L^2}{\kappa^2 \gamma^{2}}.
\end{equation}
 For this background with nonzero chemical potential, electrical conductivity is given by
$\sigma = \frac{(sT)^2}{(\epsilon + P)^{2}} \frac{1}{g^{2}}$. Using this result we can find out thermal conductivity. On evaluating the ratio $\frac{\kappa_T\mu^{2}}{\eta T}$ one finds it to be equal to $4 \pi^{2}\gamma^{2} $. Up on evaluating the ratio $\frac{d^2}{d-2}\Big(\frac{c^{'}}{k^{'}}\Big)$ we get the same result (note that here d=3).

\end{itemize}

\begin{itemize}
 \item \textbf{  AdS/QCD ($D3/D7$):} In\cite{Ge:2008ak}, authors considered $N_c$ $D3$-branes 
and $N_f$ $D7$-branes, and treated the $D3$-branes as a gravitational background (RN-AdS black hole). This corresponds to the ${\cal N}=4$ SYM 
in finite temperature with finite baryon density. To identify $ c^{'}$ and $k^{'}$, consider limit $\mu\rightarrow 0 $. In this limit various thermodynamical quantities are given by 
\begin{eqnarray}
T 
&=&
\frac{r_{+}}{\pi l^{2}},
\\
P
&=&
\frac{l^3 \pi^{4}}{2\kappa^2} T^{4}, 
\label{pressure} 
\\
\chi=\frac{\rho}{\mu}
&=&
\frac{2 \pi^{2}l}{e^{2}}T^{2}.
\label{chi}
\end{eqnarray} Now comparing with  $P=c^{'} T^{4}$, $\frac{\rho}{\mu}=\chi= k^{'} T^{2}$, we find $c^{'}= \frac{\pi^{4}}{2}\frac{l^{3}}{K^{2}}$ and $k^{'}=2 \pi^{2}\frac{l}{e^2}$, where $l$ is AdS radius, $ K $ and $e$ are gravitational and electromagnetic coupling constants respectively. Ratio of $e$ and $K$ can be expressed in terms $N_c$ and $N_f$ as $\frac{e^{2}}{K^{2}}=\frac{N_{c}}{N_{f}}l^{-2}$.
 So using these values we get $\frac{\kappa_T\mu^{2}}{\eta T}=2 \pi^{2}\frac{e^{2} l^{2}}{K^{2}} = 2  \pi^{2}\frac{N_{c}}{N_{f}}$. This is exactly the result which was reported in \cite{Ge:2008ak} where authors computed electrical conductivity to find out thermal conductivity and then the required ratio.
\end{itemize}

\begin{itemize}
 \item {\textbf{Five dimensional R-charged black hole:}}
Here we collect all the relevant information for five dimensional R-charged black hole. The Lagrangian is
\begin{equation}
 {{\cal L}\over
 \sqrt{-g} }=
 R  - {1\over 4} G_{ij} F_{\mu\nu}^i 
 F^{\mu \nu\, j} +.....,
\label{lagrangian}
\end{equation}
where
$$
G_{ij} = {L^{2}\over 2} \mbox{diag} \left[ (X^1)^{-2}, \, (X^2)^{-2}, \,(X^3)^{-2}
\right]\,.
$$
Metric and gauge fields are (for single R-charge)
\begin{equation}
ds^2_5 = - { H}^{-2/3}{(\pi T_0 L)^2 \over u}\,f \, dt^2 
+   H^{1/3}{(\pi T_0 L)^2 \over u}\, \left( dx^2 + dy^2 + dz^2\right)
+ H^{1/3}{L^2 \over 4 f u^2} du^2\,,
\label{metric_u_3}
\end{equation}
\begin{equation}
f(u) =  H(u) - u^2  (1+\kappa_1)\,, 
\;\;\;\;\; H = 1 + \kappa_1 u \,, \;\;\;\;\; 
\label{identif}
\end{equation}
\begin{equation}
A_t = { \pi T_0 \sqrt{2k_1(1+k_1)} u\over  H(u)}\,,\qquad  
\label{scal_gauge_u_3}
\end{equation}
\begin{equation}
 \frac{1}{16\pi G_{5}} =\frac{N^{2}}{16 \pi^{2}L^{3}}.
\end{equation}
Viscosity and various thermodynamical quantities are given by
\begin{equation}
T = 
{2 + \kappa_1 \over 
2\sqrt{(1+\kappa_1)}}\, T_0\,\label{temp}, 
\end{equation} 
\begin{equation}
\eta =  {\pi N^2 T^3 \over 8}\frac{(1+\kappa_1)^{2}}{(1+\frac{\kappa_1}{2})^2}  \, , 
\label{eta}
\end{equation}
\begin{equation}  P =  {\pi^2 N^2 T^4 \over 8}\frac{(1+\kappa_1)^{3}}{(1+\frac{\kappa_1}{2})^4}\,. 
\label{pressure}
\end{equation} where $T_0$ is the temperature at vanishing $\kappa_1$ i.e. at vanishing chemical potential.
The charge density is given by 
\begin{equation}
\rho = 
{\pi N^2 T_0^3\over 8} \sqrt{2 \kappa_1}(1+\kappa_1)^{1/2}\,.  
\end{equation}
The chemical potential conjugate to $\rho$ is defined as
\begin{equation}
\mu= A_t (u)\Biggl|_{u=1} = \frac{\pi T_0 \sqrt{2 \kappa_1}} {(1+\kappa_1)}
 (1+\kappa_1)^{1/2}\,,  
\end{equation} so that susceptibility is given by
\begin{equation}
\chi = \frac{\rho}{\mu} ={ N^2 T^2\over 8} \frac{(1+\kappa_1)^{2}}{(1+\frac{\kappa_1}{2})^2} ,\label{chi}
\end{equation} where we have used eqn. (\ref{temp}) to express $T_0$ in terms of T.

Upon comparing  eqn.(\ref{P}) and eqn.(\ref{et}) with eqn.(\ref{eta}), eqn.(\ref{pressure})  and eqn.(\ref{chi})  we get
\begin{equation}
 f_{\chi}(m)= \frac{ (1 + \kappa_1)^2}{(1 + \frac{\kappa_1}{2} )^{2}},~~~~f_{\eta}(m)=\frac{ (1 + \kappa_1)^2 }{(1 + \frac{\kappa_1}{2} )^3},~~~~
f_{p}(m)= \frac{(1 + \kappa_1)^3 }{(1 + \frac{\kappa_1}{2} )^4},~~c^{'}= \frac{\pi^2 N^2}{8},~~k^{'}=\frac{N^{2}}{8} \label{functions}.
\end{equation}
\end{itemize}
\begin{itemize}
 \item {\textbf{4 and 7 dimensional R-charge black holes:}}
 In order to avoid repetition, here we just list values of $c^{'}$ and $k^{'}$ which were used in the Table.  
In four dimensions we have $k^{'}=\frac{N_c^{\frac{3}{2}}}{18\sqrt{2}}$, and 
 $c^{'}=\frac{\sqrt{2}\, \pi^2}{3}\, \left( \frac{2}{3} \right)^3\,N_c^{3/2}$.
 In seven dimensions we have $k^{'}= \left( \frac{2}{3} \right)^5 N_{c}^3 \pi$, and
$c^{'}= \frac{\pi^3}{2}\, \left( \frac{2}{3} \right)^7\, N_c^3$. See \cite{Jain:2009pw,Maeda:2008hn,Behrndt:1998jd,Jain:2009uj,Gubser:1998jb} for details of the geometry.
\end{itemize}

\end{document}